\documentclass[12pt]{article}

\usepackage{graphicx}

\begin{document}

\input{epsf.sty}

\begin{titlepage}

\begin{flushright}
IUHET-486\\
quant-ph/0507252
\end{flushright}
\vskip 2.5cm

\begin{center}
{\Large \bf Nonunitary Quantum Theory with a Field Cutoff}
\end{center}

\vspace{1ex}

\begin{center}
{\large B. Altschul\footnote{{\tt baltschu@indiana.edu}}}

\vspace{5mm}
{\sl Department of Physics} \\
{\sl Indiana University} \\
{\sl Bloomington, IN 47405 USA} \\

\end{center}

\vspace{2.5ex}

\medskip

\centerline {\bf Abstract}

\bigskip

We consider a scalar quantum field theory, in which the interaction takes the form
of a field cutoff; the energy diverges to infinity whenever the value of the
field at some point falls outside a finite interval. In a simple
(1+1)-dimensional version of this theory, we may calculate the results of certain
scattering processes exactly. The main feature of the nontrivial solutions is
the appearance of shock fronts, whose time development is irreversible. The
resulting nonunitarity implies that these
theories are, at a minimum, radically different from conventional quantum field
theories.

\bigskip

\end{titlepage}

\newpage

Recently, there has been interest in quantum field theories with field cutoffs.
In these theories, values of quantized bosonic fields above some particular
cutoff value
are excluded. These kinds of theories could be interesting in their own right,
or they might be merely computational devices. In any case, the physics of
systems with field cutoffs is not very well understood, and further analysis of
these kinds of interactions is warranted.

Our inquiry into theories with field cutoffs is motivated from several directions.
One motivation is very basic; the question of whether these cutoffs are allowed
and interesting is a fairly elementary one.
In nonrelativistic quantum mechanics, problems with infinite potential barriers
are studied quite frequently. In these situations, when position is the quantized
variable, this kind of cutoff is not pathological. In fact, a cutoff in position
space usually results in a simpler, more tractable theory. We would like to know
whether this can be generalized to more realistic theories. Relativistic
single-particle theories have many attendant problems, so the most natural
generalization of these nonrelativistic ideas is directly to relativistic quantum
field theory.

Interactions with field cutoffs have already been discussed in studies of the
renormalization
group (RG). They could arise as a result of nonlinearities in the RG
transformation~\cite{ref-morris} or in the limit in which the number of quantum
fields goes to infinity~\cite{ref-gies}. The cutoff arises from the fact that
the potential diverges for values of the field outside some particular range.
However, there appears to be disagreement in the literature as to whether this
cutoff behavior makes a theory inherently pathological. This is a major question
that we shall seek to answer.

This sort of theory also arises in condensed matter contexts. Models with field
cutoffs have been introduced to study the dynamics of membrane
stacks~\cite{ref-helfrich,ref-janke1,ref-janke2,ref-leibler,ref-sornette} and the
charge stripe states of
superconductors~\cite{ref-zaanen}. These systems contain arrays of extended
objects which can undergo deformations; however, the individual membranes or
stripes cannot
cross one-another. If the quantized field describes the deformation of a single
sheet, then the field cutoff is an approximation to this no crossing condition.
A number of different workers have recently
studied the properties of these models using a variety of techniques.
The analytical results of~\cite{ref-zaanen} are in fairly good agreement with
those found in~\cite{ref-nishiyama1,ref-nishiyama2} using numerical density matrix
RG methods. These papers emphasize the fact that the cutoff transforms
short-wavelength fluctuations into longer-wavelength ones.
However, their conclusions disagree with those of of~\cite{ref-orland}.
We shall look at these theories from a very different viewpoint than that which
is taken in these papers; however, a better
understanding of the general features of these models may help to resolve
this controversy.

Field cutoffs have also been introduced as purely computational devices. By
eliminating large field values from the measure of a path integral, one may
generate a modified perturbation series with quite good convergence
properties~\cite{ref-pernice,ref-meurice1}. This technique has been profitably
applied to the anharmonic oscillator~\cite{ref-li1} and has also been adapted for
use in lattice gauge theory~\cite{ref-li2,ref-li3}. Some work has also been
done on finding the optimal value of the field cutoff for computational
purposes~\cite{ref-li3,ref-kessler}. While using a field cutoff in this
way is very interesting, it will not be directly relevant here. 
%However, our analysis might be valuable for
%future examinations of the usefulness of a field cutoff as an approximation
%technique.

We want to consider a theory in which the field cutoff is a real physical
effect---i.e., an interaction.
We may describe such a theory schematically by the Lagrange density
\begin{equation}
{\cal L}={\cal L}_{0}-V(\phi)=\frac{1}{2}(\partial^{\mu}\phi)(\partial_{\mu}\phi)
-\frac{m^{2}}{2}\phi^{2}-V(\phi).
\end{equation}
${\cal L}_{0}$ describes free propagation of plane waves, and the potential
$V(\phi)$ parameterizes the interaction. The key property of $V(\phi)$ is that
it must diverge to
positive infinity for all values of the field $|\phi|>\Phi$. It would
require an infinite amount of energy to push the value of the field above this
limit at even a single point.

If the potential contains no additional self interactions beyond the cutoff
itself, then $V$ can be roughly seen as the $n\rightarrow\infty$ limit of
the sequence of potentials
\begin{equation}
U_{n}(\phi)=\lambda\mu^{d}\left(\frac{\phi^{2}}{\Phi^{2}}\right)^{n}
\end{equation}
(where $d$ is the dimensionality and $\mu$ is an additional mass scale inserted to
leave
$\lambda$ dimensionless). For large enough values of $n$ and $d>2$,
these interactions will
be nonrenormalizable. In the language of effective field theory, the interactions
are irrelevant. However, a RG analysis is not
useful for the $n\rightarrow\infty$ limit. In this limit, no matter how small the
coupling
$\lambda$ becomes, the potential still diverges as $\phi\rightarrow\Phi$, so the
physics remains unchanged, despite the apparent irrelevance of the interaction.

Standard methods for calculations in quantum field theory therefore are not
useful. A different approach is needed. We shall perform an exact calculation for
a scattering process in a particularly simple version of this kind of theory.
We shall use methods from relativistic quantum mechanics, looking at the evolution
of particular field profiles. This may be translated into the language of
quantum field theory through the use of a wave functional $\Psi[\phi(x),t]$,
which gives the amplitude for the system to be found with a particular functional
form $\phi(x)$ for the field at a time $t$.

Our method is based upon the idea that, when the field $\phi$ reaches its cutoff
value $\Phi$ at some point, then we must
effectively introduce a new boundary condition at that
point. This condition will account for the fact that no flux can pass through
this boundary, because doing so would increase $\phi$ to a value greater than
$\Phi$, which would result in infinite energy. (We consider values $\phi>\Phi$ to
be forbidden, but the crossover point in field space, $\phi=\Phi$, is allowed.)
We may then examine the physics of this new boundary, which will be determined
from conservation laws.

The results we shall find are quite interesting. The system with the field
cutoff can describe a perfectly consistent wave propagation scheme that is
calculationally tractable. The key feature is the development of shock fronts,
whose propagation is governed by a new evolution equation.
However, the resulting time development is nonunitary
and thus decidedly different from that found in conventional quantum theory. It
is therefore unclear whether theories of this type can be considered viable as
models for any real-world phenomena.

We shall now consider our specific example, illustrating how this theory behaves.
For simplicity, we work in 1+1 dimensions and with a massless theory. $V$ has
the form of a pure cutoff function,
\begin{equation}
V(\phi)=\left\{
\begin{array}{ll}
0, & |\phi| < \Phi \\
+\infty, & |\phi| > \Phi
\end{array}
\right..
\end{equation}
Our wave packets will also have a particularly simple triangular form. At large
negative times, the field configuration is
\begin{equation}
\label{eq-phii}
\phi_{i}(x,t)=f_{i}^{+}(x-t)+f_{i}^{-}(x+t),
\end{equation}
where $f_{i}^{+}$ and $f_{i}^{-}$ are the initial right- and left-moving wave
packets. Neither contains any point where the field value exceeds $\Phi$, so
they propagate according to the free massless Klein-Gordon equation---i.e., the
wave equation, $\left(\partial_{t}^{2}-\partial_{x}^{2}\right)\phi=0$. The
two incoming packets are identical in functional form:
\begin{equation}
f_{i}^{\pm}(x)=\left\{
\begin{array}{lc}
\frac{3}{4}\Phi\left(1+\frac{x}{w}\right), & -w\leq x\leq0 \\
\frac{3}{4}\Phi\left(1-\frac{x}{w}\right), & 0\leq x\leq w \\
0, & |x|\geq w
\end{array}
\right..
\end{equation}
The total width of each wave packet is $2w$, and the full width at half maximum
is $w$. These wave packets are not normalized in any way, but the Klein-Gordon
theory has no probabilistic interpretation. A real-valued wave packet propagating
in one
direction contains equal parts positive- and negative-energy plane waves, and
the conventional current, proportional to $\Im\left\{\phi^{*}\partial^{\mu}\phi
\right\}$,
obviously vanishes. What is conserved in the free
evolution of these wave packets is generally the integral of $\phi(x,t)$ over
all space, and we shall make important use of this conservation law in our
calculation.

It is clear that when the wave packets get close together, around $t\approx0$,
the field will reach its cutoff value at some point.
In fact, This first occurs at time $t=-\frac{w}{3}$, when the field
profile is
\begin{equation}
\phi\left(x,-\frac{w}{3}\right)=\left\{
\begin{array}{lc}
\frac{3}{4}\Phi\left(\frac{4}{3}+\frac{x}{w}\right), & -\frac{4w}{3}\leq x\leq
-\frac{2w}{3} \\
\frac{3}{2}\Phi\left(1+\frac{x}{w}\right), & -\frac{2w}{3}\leq x\leq
-\frac{w}{3} \\
\Phi, & -\frac{w}{3}\leq x\leq \frac{w}{3}w\\
\frac{3}{2}\Phi\left(1+\frac{x}{w}\right), & \frac{w}{3}\leq x\leq
\frac{2w}{3} \\
\frac{3}{4}\Phi\left(\frac{4}{3}-\frac{x}{w}\right), & \frac{2w}{3}\leq x\leq
\frac{4w}{3} \\
0, & |x|\geq\frac{4}{3}w
\end{array}
\right..
\end{equation}
It is a quirk of triangular wave packets that at every point in a
finite-width region the field attains the value $\Phi$ simultaneously.
More typically, the cutoff will be reached first at a single point; then that
field value will spread outward to cover a region on nonzero width. However,
despite the special situation that exists here, the subsequent evolution
does not seem to differ that markedly from the
generic case; the aforementioned spreading of the $\phi=\Phi$ region will occur
in this instance as well.

By the time the wave packets interpenetrate enough for the cutoff value to be
reached, the leading edges of the packets have progressed beyond the cutoff
region. Portions that are far enough along will continue to propagate freely
and will escape to infinity without any interruptions of their motions. (We shall
refer to these unmodified portions of the two original wave packets as the
outgoing parts, since they have already progressed beyond the region where
interactions are present. Similarly, the portions of the wave packet that have not
yet reached the cutoff region will be referred to as the incoming parts.)

However, the fate of that portion of the field lying within the cutoff region
is very different.
The wave fronts will remain stationary wherever $\phi=\Phi$.
On the surface, this might seem inconsistent with the fact that the kinetic
part of the Lagrangian implies purely lightlike propagation. However, since
the interaction has an infinite strength, it dominates whenever it is nonzero;
the kinetic parts of ${\cal L}$ are entirely negligible in comparison. Normal,
free propagation effects are suppressed. There is also another way to look at
the resolution of this apparent paradox. Every point at which $\phi=\Phi$
is infinitely repulsive. Inside the cutoff region, every point along the field
profile is hemmed in on either side by two impenetrable walls. The field
oscillates back and forth between these two walls and so remains stationary.
This is related to the fact that the local energy (i.e. frequency) is infinite,
so we expect such infinitely rapid oscillations.

The interior of the cutoff region is therefore locked in place. However, at the
edges, where the potential is discontinuous, this argument would not necessarily
hold. There would appear to be a
reflecting boundary on the inward side only, so that the field could flow away
on the outward side. Ultimately, this is the mechanism by which the region will
decay. However, the decay will not begin immediately; on the contrary, the
region will continue to expand for some time, as incoming flux builds up
against the region's edges.

At $t=-\frac{w}{3}$, waves are still flowing in toward in toward the edge of the
cutoff region. The wave fronts cannot penetrate into the region where $\phi=\Phi$.
However, they cannot be reflected back on themselves either, because then linear
superposition would result in field values greater than $\Phi$. What must
happen instead is that each point along the field profile propagates inward
toward the interaction region until it reaches a point when it can progress no
further. At that point, it must become part of the cutoff region. (The only
alternative would be reflection, and we know that that would be impossible.)
So the $\phi=\Phi$ region will expand as the incoming waves add to it. The
region's edge will move outward as a shock. We must now determine the behavior
of this shock. (There are naturally two shocks, one with positive $x$ and one
with negative. We will concentrate our attention on the positive side; the
characteristics of the other are simply related by parity.)

Let $y(\tau)$ be the distance the shock has advanced in the time $\tau$ since its
formation at $x=\frac{w}{3}$. That is, $x_{s}=y+\frac{w}{3}$ is the position of
the shock front at a time $t=\tau-\frac{w}{3}$. We shall determine $y$ from a
self-consistency condition. The shock swallows up wave material as it moves,
and the total material that it has absorbed determines the size of the region
behind it. The ``volume'' (in the two-dimensional ``length$\times$field space'')
of the region behind the shock (excluding that part which existed at $\tau=0$)
is simply $y\Phi$. This volume must be made up of wave material that has been
caught behind the shock. There are two contributions to this volume, which we
shall denote by $y_{1}\Phi$ and $y_{2}\Phi$. The remaining incoming wave
contributes to $y_{1}$ and the outgoing wave to $y_{2}$.

The incoming wave will build material up behind the shock as it flows past the
shock front. The flux from the incoming wave will
slow with time, as the incoming amplitude at the shock front diminishes. The total
volume that has propagated to the left of the point $y$ between the
formation of the shock and time $\tau$ is
\begin{equation}
y_{1}\Phi=\int_{0}^{\min(y+\tau,w)}dy'\,\frac{3}{4}\Phi\left(1-\frac{y'}{w}
\right).
\end{equation}
The integrand is just the initial profile of the incoming wave packet. The upper
limit on the integrand is either $w$ (if the entire packet has been absorbed) or
$y+\tau$. The $+\tau$ comes from the fact that the wave is moving to the left;
even if $y$ were somehow held constant, elements of the wave packet
would still be flowing past
the shock with speed $v=1$, and so the volume would continue to increase.

The second contribution to $y$ is slightly trickier. It comes from the amount
of material in the outgoing wave packet that is captured by the shock. Although
the outgoing packet is moving at the speed of light, the shock can actually
advance superluminally. So long as the shock speed remains greater than unity,
the contribution from the outgoing wave is
\begin{equation}
y_{2}\Phi=\int_{0}^{\min\left(y-\tau,\frac{w}{3}\right)}dy''\,\frac{1}{4}\Phi
\left(1-\frac{3y''}{w}\right).
\end{equation}
The sign of $\tau$ is opposite what it was in $y_{1}$, because the wave 
being absorbed is
moving in the opposite direction; in this case, if $y$ were held constant, there
would be zero contribution to $y_{2}$, because the outgoing wave would simply
move away from the shock front unhindered.

For small enough $\tau$, $y=y_{1}+y_{2}$ is given by
\begin{eqnarray}
y & = & \frac{3}{4}(y+\tau)-\frac{3}{8w}(y+\tau)^{2}+\frac{1}{4}(y-\tau)-
\frac{3}{8w}(y-\tau)^{2} \\
& = & \sqrt{\frac{2w\tau}{3}-\tau^{2}}.
\end{eqnarray}
We can easily verify that the shock speed is initially greater than one. In
fact, $\dot{y}(\tau=0)$ diverges. However, $\dot{y}$ decreases rapidly after
that point, until it reaches the value $\dot{y}=1$ at $\tau=\frac{w}{3}-\frac{w}
{3\sqrt{2}}$. The position of the shock at that time is $y=\frac{w}{3\sqrt{2}}$.
After this time, the shock wave has slowed sufficiently that the portion of
the outgoing wave packet that remains unabsorbed will be able to escape to
infinity, because that packet is now advancing more quickly than the shock.

The superluminal shock propagation may seem problematic. There is clearly some
sort of microcausality violation occurring. The existence of such a violation
should not be too surprising, since the Lagrangian for this theory has a
singular structure. In light of this fact, one might be inclined to dismiss the
theory as unworkable. However, this is not necessary, because the theory
maintains a macroscopic causality. This fact follows from one relatively
simple observation. Every component of the field profile is at any given time
either moving at the speed of light or stationary (if it lies in a region where
$\phi$ is at its cutoff value). Under these circumstances, a given element of
the profile can never propagate to a location outside its future light cone. This
means, for example, that the leading edge of a packet can never cross a distance
$d$ in a time less than $d$.

Fundamentally, while the free Lagrange density ${\cal L}_{0}=\frac{1}{2}
(\partial^{\mu}\phi)(\partial_{\mu}\phi)$ ensures microcausal propagation,
the interaction part, which always dominates when it is nonzero, does not have
a manifestly causal structure. As long as propagation is governed by
${\cal L}_{0}$ only, causality is guaranteed, but wherever $\phi=\Phi$, the
physics are entirely different, and it is exactly the dynamics of these cutoff
regions that display the causality difficulties.

Let us now return to the time evolution of the colliding wave packets. The
cutoff region has expanded, because incoming waves with amplitudes greater than
$\frac{1}{2}\Phi$ cannot be reflected back to infinity. This effect
governs the evolution up to the point at which $y+\tau=x_{s}+t=\frac{w}{3}$. This
is precisely when the amplitude of the incoming wave falls to $\frac{1}{2}\Phi$.
The incoming waves can then be reflected without generating any field values
$\phi\geq\Phi$. So no further material will be built up at the shock front.
Furthermore,
the cutoff region must immediately begin to decay, since without the accumulation
of material at its edges, there are no longer any hard walls confining it.

Let us ignore for a moment the surviving pieces of the original incoming wave
packets.
The cutoff region contributes
\begin{equation}
\phi_{c}\left(x,t=-\frac{w}{3\sqrt{2}}\right)=\Phi\left[\theta\left(x+\frac{w}{3}+
\frac{w}{3\sqrt{2}}\right)-\theta\left(x-\frac{w}{3}-\frac{w}{3\sqrt{2}}\right)
\right]
\end{equation}
to the field at the time $t=-\frac{w}{3\sqrt{2}}$. The time derivative of
$\phi_{c}$ is zero everywhere except at the boundaries. After this time, flux can
travel out from the boundaries in either direction. Solving the wave equation
with these initial conditions, we find that for $t\geq-\frac{w}{3\sqrt{2}}$, the
configuration evolves into
\begin{eqnarray}
\phi_{c}(x,t)& = & \frac{1}{2}\Phi\left[
\theta\left(x-t+\frac{w}{3}\right)
-\theta\left(x-t-\frac{w}{3}-\frac{w\sqrt{2}}{3}\right)\right. \nonumber\\
& & +\left.\theta\left(x+t+\frac{w}{3}+\frac{w\sqrt{2}}{3}\right)
-\theta\left(x+t-\frac{w}{3}\right)\right].
\end{eqnarray}
This consists of equal right- and left-moving parts. The flux begins to exit
through the shock
in either direction, and as it does, the size of the cutoff region diminishes.
The hard walls recede toward $x=0$ with unit velocity. This keeps the edges of
the shocks just ahead of the fronts of the remaining incoming packets. The
remains of the incoming packets never reach the barriers and so are therefore
not reflected. In fact, the whole system now reverts to free evolution
under the wave equation.

\begin{figure}[t]
\epsfxsize=3in
\begin{center}
\leavevmode
\epsfbox{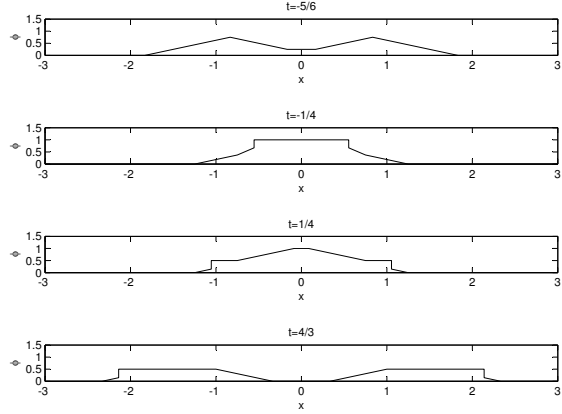}
\caption{The field at four points in time, with variables in dimensionless form.
The position $x$ and time $t$ are given in units
of $w$, while $\phi$ is in units of the cutoff field $\Phi$. At $t=-5/6$, the
two initial packets are just beginning to interpenetrate; no interaction has yet
occurred. By $t=-1/4$, the shock fronts have come into existence and are
expanding. At $t=1/4$, the shocks are receding again, and at $t=4/3$, the two
final wave packets have separated completely.
\label{fig-plots}}
\end{center}
\end{figure}

The final wave packets are
\begin{eqnarray}
f_{f}^{+}(x) & = & \left\{
\begin{array}{lc}
\frac{3}{4}\Phi\left(1+\frac{x}{w}\right), & -w\leq x\leq-\frac{w}{3} \\
\frac{1}{2}\Phi, & -\frac{w}{3}\leq x < \frac{w}{3}+\frac{w\sqrt{2}}{3} \\
\frac{3}{4}\Phi\left(1-\frac{x}{w}\right), &  \frac{w}{3}+\frac{w\sqrt{2}}{3} <
x\leq w \\
0, & |x|\geq w
\end{array}
\right. \\
f_{f}^{-}(x) & = & f_{f}^{+}(-x),
\end{eqnarray}
with
\begin{equation}
\label{eq-phif}
\phi_{f}(x,t)=f_{f}^{+}(x-t)+f_{f}^{-}(x+t)
\end{equation}
for times $t\geq-\frac{w}{3\sqrt{2}}$. The incoming expression (\ref{eq-phii}) was
valid for all times $t\leq-\frac{w}{3}$, and at intermediate times, the field
configuration is
\begin{equation}
\phi(x,t)=\left\{
\begin{array}{lc}
\Phi, & |x|<\frac{w}{3}+\sqrt{\left(\frac{w}{3}\right)^{2}-t^{2}} \\
f_{i}^{+}(x-t)+f_{i}^{-}(x+t), & |x|>\frac{w}{3}+\sqrt{\left(\frac{w}{3}
\right)^{2}-t^{2}}
\end{array}
\right..
\end{equation}The field profiles, at four representative times, are displayed in
Figure~\ref{fig-plots}. Note that the interaction has flattened out the final
packets, and many short-distance features have been lost.

%\begin{figure}[t]
%\begin{center}
%\includegraphics[height=0.4\textwidth,angle=0]{cutoffplots.pdf}
%\caption{The field at four points in time, with variables in dimensionless form.
%The position $x$ and time $t$ are given in units
%of $w$, while $\phi$ is in units of the cutoff field $\Phi$. At $t=-5/6$, the
%two initial packets are just beginning to interpenetrate; no interaction has yet
%occurred. By $t=-1/4$, the shock fronts have come into existence and are
%expanding. At $t=1/4$, the shocks are receding again, and at $t=4/3$, the two
%final wave packets have separated completely.
%\label{fig-plots}}
%\end{center}
%\end{figure}

Finally, we may determine the shock positions at all times. (The shocks
continue to persist, moving backwards, even after the free evolution resumes.
The fact that the evolution is free only implies that the shocks must move with
unit velocity.) The shocks form at time $t=-\frac{w}{3}$ and positions
$x_{s}\left(
-\frac{w}{3}\right)=\pm\frac{w}{3}$. Their subsequent evolution is
\begin{equation}
x_{s}=\pm\left\{
\begin{array}{lc}
\frac{w}{3}+\sqrt{\left(\frac{w}{3}\right)^{2}-t^{2}}, & -\frac{w}{3}\leq t\leq
-\frac{w}{3\sqrt{2}} \\
\frac{w}{3}-t, & -\frac{w}{3\sqrt{2}}\leq t < \frac{w}{3}
\end{array}
\right..
\end{equation}
At time $t=\frac{w}{3}$, the shocks reach one-another at $x_{s}=0$ and cease to
exist.

Note that $f_{i}^{\pm}$ and $f_{f}^{\pm}$ agree at every spacetime point that is
outside the future light cones of all the $\phi=\Phi$ points. This is an
illustration of
the fact that if the worldline of a point does not encounter the cutoff region,
it will propagate freely forever. In general, we expect that whenever two wave
packets collide, their front edges will pass through one-another and emerge
essentially unmodified. The trailing edges should behave similarly, because
the cutoff region will begin to dissipate before those parts of the packet can
reach it.

The interaction we have considered is microscopically repulsive, and the results
we have found are in keeping with this fact. The centroid of $f_{f}^{+}(x)$ lies
at $x=\frac{w}{27}+\frac{2w\sqrt{2}}{81}$, while the centroid of $f_{i}^{+}(x)$
is obviously at $x=0$. So the final right-moving wave packet is advanced
relative to the initial right-moving packet. When the field components
are moving, they always move with the speed of light. This means that the final
packets must consist primarily of reflected, rather than transmitted, waves. For,
if the wave were primarily transmitted, the positions of the centroids would imply
superluminal propagation.
Roughly speaking, the incoming waves are partially reflected by
the shock and rebound back towards the directions from which they originated.
However, this picture cannot be made much more precise, because of the quantum
indistinguishability of the field amplitudes.

We may generalize our calculations somewhat. If we initially have two
colliding wave packets, described by continuous functions $\bar{f}_{i}^{+}(x-t)$
and $\bar{f}_{i}^{-}(x+t)$ that
are mirror images of one-another---$\bar{f}_{i}^{+}(x)=\bar{f}_{i}^{-}(-x)$---and
have only a single peak each (i.e. the packets are first monotonically increasing,
then decreasing), then the final wave packets will have a particular form. They
will clearly remain mirror images, and the right-moving packet will be
\begin{equation}
\bar{f}_{f}^{+}(x)=\left\{
\begin{array}{lc}
\bar{f}_{i}^{+}(x), & x < x_{1} \\ 
\frac{1}{2}\Phi, & x_{1} < x < x_{2} \\
\bar{f}_{i}^{+}(x), & x > x_{2} 
\end{array}
\right..
\end{equation}
The points $x_{1}$ and $x_{2}$ are determined by simple conditions.
$x_{1}$ is the minimum value at which $\bar{f}_{i}^{+}(x)$ is equal to
$\frac{1}{2}$, and $\int_{x_{1}}^{x_{2}}dx\, \bar{f}_{i}^{+}(x)=\frac{1}{2}$.

The only nontrivial aspect of the time evolution we have studied
is the shock propagation.
This has actually been a fairly typical shock evolution problem. The shock
appears at a point in time when the weakly interacting equations of motion lose
their validity. Then the time development of the shock front is governed by
conservation laws. These are both quite standard features. Shocks are generally
difficult to treat analytically, unless there are some significant simplifying
features. We utilized three such features---low dimensionality, dispersionless
propagation, and particularly
simple wave packets---in our example.
However, there are many computational techniques that apply
in more general situations, and these methods should also apply to more
complicated quantum theories with field cutoffs.

One other feature that is typical of shock problems is irreversibility, and
that occurs in this system as well. Depending on one's point of view, this could
either make the theory highly interesting or potentially exclude it completely.
At the very least, it shows that there is a marked difference between this
system and more conventional quantum systems.

To see the irreversibility explicitly, we observe that if the incoming wave
packets, for $t\ll 0$, are given by (\ref{eq-phif}), then (\ref{eq-phif})
actually remains valid at all times. Shock fronts will come into existence, and
their evolution may be determined by the same methods we used previously.
However, in this case, the appearance of the shocks will not change the time
evolution in any way. This is tied to the fact that the free evolution does not
lead to there being any point with a field value strictly greater than $\Phi$.

In the language of the second quantized Klein-Gordon theory, the irreversibility
means the evolution is nonunitary. Consider a wave functional $\Psi[\phi(x),t)]$
which is a superposition of two different states at a large negative time $-T$:
\begin{equation}
\Psi[\phi_{i}(x,-T),-T]=\Psi[\phi_{f}(x,-T),-T]=\frac{1}{\sqrt{2}}.
\end{equation}
We now
take $\phi_{i}$ and $\phi_{f}$ to be defined by (\ref{eq-phii}) and
(\ref{eq-phif}) at all times. The time $-T$ is a parameter when it appears
as an argument in $\phi$, but it is the dynamical time coordinate when it is
an argument of $\Psi$.
Both of the incoming states evolve into the same outgoing state, so for
large times $\Psi[\phi_{f}(x,T),T]=\sqrt{2}$. The normalization of the wave
functional has changed. This makes the theory inconsistent with a conventional
probabilistic interpretation.

However, a modified probabilistic interpretation may be possible. If the initial
conditions contain only well localized wave packets, then at large enough times,
the field profile can still only contain
freely evolving wave packets. If we wait for a time large enough that all
possible scattering events are complete, we may then adjust the normalization of
the wave functional to give a new probabilistic interpretation. This method is
{\em ad hoc}, and it would not be expected to give meaningful results when the
wave packet interactions are still going on. However, renormalized interacting
quantum field theories do not have valid probabilistic interpretations during the
interaction period either. These theories require regulation, which means summing
over a set of intermediate states that is truncated. Without a complete set of
intermediate states, probability conservation will generally fail. A sensible
probabilistic interpretation can exist only for the asymptotic states. (Of course,
the problems with unitarity in a renormalized quantum field theory are still
much less profound than in the case presently under consideration.)

If a field cutoff were introduced into the theory by some artificial means (say,
as a calculational device), then a breakdown of unitarity would not necessarily be
unexpected. We might conclude that the breakdown was merely a consequence of
the approximations that the field cutoff entailed. For example, if a model with
a cutoff
were used to describe the dynamics of a single membrane in a stack, we would
expect that a more detailed model, in which the dynamics of all the membranes
were considered, would preserve unitarity.
However, no such argument can be made if a diverging potential is part of the
fundamental theory. The field cutoff then comes directly from the dynamics, and no
approximations are involved.

The nonunitarity of the time evolution operator is a consequence of the
fact that this theory does not have a well-defined
Hermitian Hamiltonian operator. The
energy exhibits an infinite discontinuity as $\phi\rightarrow\Phi$ at any point.
There is certainly interest in modified quantum theories with nonunitary
evolution. Much speculation has focused on the possibility that gravity, either
through quantum gravity or cosmological effects, may be the source of the
nonunitarity. (See, for example~\cite{ref-milburn,ref-gambini,ref-calucci}.) This
kind of theory, with a field cutoff, provides a completely
different microscopic explanation of how irreversibility occurs. Moreover,
while gravity-induced nonunitarity typically involves decoherence---i.e., a
superposition state evolving into a mixed state---we have demonstrated
nonunitarity with increased coherence, with the norm of the wave functional
increasing rather than decreasing.

We have certainly not considered the most general theory in which a field cutoff
exists. It is possible that including further complexities, such as a source of
dispersion in the free propagation (e.g., a mass term) or additional interactions
beyond the field cutoff itself, could change the main result that we have
obtained. Unfortunately, studying shock propagation in the presence of significant
dispersion can be quite difficult. Similarly, including, say, a $\phi^{4}$
interaction
in $V$ would complicate the theory tremendously. In either of these situations,
it might be extremely difficult to obtain reliable results.

However, there is no strong reason to believe that the nonunitarity we have
observed does not carry over to more complicated cases. Our ultimate result is
that, in one highly suggestive special case, a quantum field theory with
a field cutoff displays irreversible time evolution. This makes the theory
profoundly different from those usually encountered and could be cause for
rejecting such a theory altogether. If theories with field cutoffs are
encountered in nonperturbative RG calculations, there is significant cause to
believe that they may be unphysical.
If such theories are not rejected outright, they must
be examined very carefully, for further pitfalls could well exist. This
therefore represents a potentially interesting line for future research.

\section*{Acknowledgments}
The author is grateful to V. A. Kosteleck\'{y} for helpful discussions and to
P. Orland for pointing out several useful references.
This work is supported in part by funds provided by the U. S.
Department of Energy (D.O.E.) under cooperative research agreement
DE-FG02-91ER40661.

\end{document}